\documentstyle[prd,aps]{revtex}
\begin{document}
\input epsf
\draft
\renewcommand{\topfraction}{0.8}
\twocolumn[\hsize\textwidth\columnwidth\hsize\csname
@twocolumnfalse\endcsname
\preprint{SU-ITP-98-65, hep-ph/9812289,   December 8, 1998}
\title { \Large  \bf   Instant Preheating }
\author{Gary Felder,$^1$ ~Lev Kofman,$^{2,3}$ ~and  ~Andrei
Linde$^1$
 }
\date {December 8, 1998}
\address{
${}^1$Department of Physics, Stanford University, Stanford, CA
94305-4060, USA \\
${}^2$CITA, University of Toronto, 60 St George Str,
Toronto, ON M5S 1A1, Canada\\
${}^3$ CEA/Saclay, Service de Physique Th\'eorique, F-91191 France}
\maketitle
\begin{abstract}
We describe a new efficient mechanism of reheating. Immediately  
after rolling
down the rapidly moving inflaton field $\phi$ produces  particles  
$\chi$, which
may be either bosons or fermions. This is a nonperturbative process which
occurs almost instantly; no oscillations or parametric resonance is  
required.
The effective masses of the $\chi$ particles may be very small at  
the moment
when they are produced, but they ``fatten'' when the field $\phi$  
increases.
When the  particles $\chi$ become sufficiently heavy, they rapidly  
decay to
other, lighter particles. This leads to an almost instantaneous reheating
accompanied by the production of particles with masses which may be  
as large as
$10^{17}-10^{18}$ GeV. This mechanism works in the usual  
inflationary models
where $V(\phi)$ has a minimum, where it takes only a half of a single
oscillation of the inflaton field $\phi$, but it is especially  
efficient  in
models with effective potentials slowly decreasing at large $\phi$  
as in the
theory of quintessence.
 \end{abstract}
\pacs{PACS: 98.80.Cq\hskip 2.3 cm SU-ITP-98-65\hskip 1cm
 CITA-98-45\hskip 2.3 cm hep-ph/9812289}
\vskip2pc]

 \section   {Introduction}~~During the last few years  we have learned
 that the first stages of reheating after
inflation are typically governed by nonperturbative effects.
In particular, the most efficient mechanism of reheating which was
known until now was based on the theory of the nonperturbative
decay of the inflaton field due to the effect of broad parametric
resonance \cite{KLS}. To distinguish  this stage of
nonperturbative particle production from the stage of particle
decay and thermalization which can be described using perturbation
theory \cite{DL} (see also \cite{new}), it was called {\it
preheating}.

This process can rapidly transfer the energy of a coherently
oscillating scalar field to the energy of other fields or
elementary particles. Because of the nonperturbative nature of the
process, it may lead to many unusual effects, such as nonthermal
cosmological phase transitions \cite{KLS96}. Another unusual
feature of preheating discovered in \cite{KLS} is the possibility
of the production of a large amount of superheavy particles  with
masses one or two orders of magnitude greater than the inflaton
mass. In the simplest versions of chaotic inflation with the
inflaton mass $m \sim 10^{13}$ GeV this can lead to the copious
production of particles with masses up to $10^{14} - 10^{15}$ GeV
\cite{KLS,Kolb,heavy,chung}. This issue is rather important since
interactions and decay of superheavy particles may lead to
baryogenesis at the GUT scale \cite{Kolb}.

However, GUT baryogenesis was only marginally possible in the
models of preheating studied until now  because the masses of
produced particles just barely approached the GUT scale. Moreover,
in some models the  particles created by the resonance strongly
interact with each other, or rapidly decay.
 This may take them out
of the resonance band, in which case the parametric resonance does
not last long or does not happen at all.  Also, there are some
models where
the effective potential does not have a minimum, but instead slowly
decreases
at large $\phi$  \cite{spok,QUINT}. In these models   the
scalar field does not oscillate at all after inflation, so neither
parametric resonance nor the standard perturbative mechanism of
inflaton decay works there.

In this paper we will try to turn these potential problems into an
advantage. We will describe a new mechanism of preheating, which
works even in the models where parametric resonance cannot
develop. The new mechanism is also nonperturbative but very
simple.  It leads to an almost instantaneous reheating accompanied
by the production of superheavy particles with masses which may be
as great as $10^{17}-10^{18}$ GeV. In some cases it may even lead
to the production of black holes of a Planckian mass, which
immediately evaporate.

\section{Instant preheating: The basic idea}

To explain the main idea of the new scenario we will   consider
the simplest model of chaotic inflation with the effective
potential $ {m^2\over 2} \phi^2$ or ${\lambda\over 4}\phi^4$ and
assume that  the inflaton field $\phi$ interacts with  some other
scalar field $\chi$   with the interaction term $-{\textstyle {1
\over 2}} g^2\phi^2\chi^2$.  In these models inflation occurs at
$|\phi| \gtrsim 0.3 M_p$ \cite{book}.  Suppose for definiteness that
initially
$\phi$ is large and negative, and inflation ends at $\phi \sim -
0.3 M_p$. After that the field $\phi$ rolls to $\phi = 0$, then it
grows up to $10^{-1} M_p \sim 10^{18}$ GeV, and finally rolls back
and oscillates about $\phi = 0$ with a gradually decreasing
amplitude. If the coupling constant $g$ is large enough ($g
\gtrsim 10^{-4}$), then, according to  \cite{KLS}, the production
of particles  $\chi$ occurs for the first time when the scalar
field $\phi$ reaches the point   $\phi = 0$ after the end of
inflation. With each subsequent oscillation, particle creation
occurs as $\phi$ crosses zero.  This mechanism of particle
production is described by the theory of preheating in the broad
resonance regime
 \cite{KLS}.  But now we   concentrate on the first instant of this
process. Remarkably, in certain cases this   is all that we need
for efficient reheating.

Usually  only a small fraction of the energy of the inflaton field $\sim
10^{-2} g^2 $ is transferred to the particles $\chi$ at that
moment (see  Eq. (\ref{ratio_0}) in the next section). The role of the
parametric resonance was to
increase this energy exponentially within several oscillations of
the inflaton field. But suppose that  the particles $\chi$
interact with fermions $\psi$ with the coupling $h \bar\psi\psi
\chi$. If this coupling is strong enough, then  $\chi$ particles
may  decay to  fermions before the oscillating field $\phi$
returns back to the minimum of the effective potential. If this
happens, parametric resonance does not occur. However, as we will
show, something   equally interesting may occur instead of it: The
energy density of the $\chi$ particles at the moment of their
decay may become much greater than their energy density at the
moment of their creation.

Indeed, prior to their decay the number density of $\chi$
particles, $n_\chi$, remains practically constant  \cite{KLS},
whereas  the effective mass of each   $\chi$ particle grows  as
$m_\chi = g \phi $ when the field $\phi$ rolls up from the minimum
of the effective potential. Therefore their total energy density
grows. One may say that $\chi$ particles are ``fattened,''  being
fed by the energy of the rolling field $\phi$. The fattened $\chi$
particles tend to decay to fermions at the moment when they have
the greatest mass, i.e. when $ \phi $ reaches its maximal value
$\sim 10^{-1} M_p$, just   before it begins rolling back to $\phi
= 0$.

At
that moment $\chi$ particles can decay to two fermions with mass
up to $m_\psi \sim {g\over 2} 10^{-1} M_p$, which can be as large
as $5\times 10^{17} $ GeV for $g \sim 1$. This is two orders of
magnitude greater than the masses of the particles which can be
produced by the usual mechanism based on parametric resonance
\cite{KLS}. As a result, the total energy density of the produced
particles also becomes two orders of magnitude greater than their
energy density at the moment of their production. Thus the chain
reaction $\phi \to \chi \to \psi$ considerably enhances the
efficiency   of transfer of energy of the inflaton field to
matter.

More importantly,  superheavy particles $\psi$    (or the products
of their decay) may eventually dominate the total energy density
of matter even if in the beginning their energy density was
relatively small. For example, the energy density of the
oscillating inflaton field in the theory with the effective
potential ${\lambda\over 4}\phi^4$ decreases as $a^{-4}$ in an
expanding universe with a scale factor $a(t)$. Meanwhile the
energy density stored in the nonrelativistic particles $\psi$
(prior to their decay) decreases only as $a^{-3}$. Therefore their
energy density rapidly becomes dominant even if originally it was
small. A subsequent decay of such particles leads to a complete
reheating of the universe.

Since the main part of the process of preheating in this scenario
(production of $\chi$ and $\psi$ particles) occurs immediately
after the end of inflation, within less than one oscillation of
the inflaton field, we will call it {\it instant preheating}. We
should emphasize that instant preheating is a completely
nonperturbative effect, which can lead to the production of
particles with momenta and masses many orders of magnitude greater
than the inflaton mass. This would be impossible in the context of
the elementary theory of reheating developed in \cite{DL}. In what
follows we will give a more detailed description of the instant
preheating  scenario.

\section{The simplest models}

Consider first the simplest model of chaotic inflation with the
effective potential    $V(\phi) = {m^2\over 2}\phi^2$,  and with
the interaction Lagrangian   $-{\textstyle {1 \over 2}}
g^2\phi^2\chi^2 - h \bar\psi\psi \chi$. We will take   $m =
10^{-6} M_p$, as required by microwave background anisotropy
\cite{book}, and in the beginning we will assume for simplicity
that $\chi$ particles do not have a bare mass, i.e. $m_\chi(\phi)
= g|\phi|$. Reheating in this model   is efficient only if $g \gtrsim
10^{-4}$ \cite{KLS,super}, which implies $g M_p \gtrsim 10^2 m$ for the
realistic value of the mass $m \sim 10^{-6} M_p$.  Thus,
immediately after the end of inflation, when $\phi \sim M_p/3$,
the effective mass $g|\phi|$ of the field $\chi$  is much greater
than $m$. It decreases when the field $\phi$ moves down, but
initially this process remains adiabatic, $|\dot m_\chi| \ll
m^2_\chi$.

The adiabaticity condition becomes violated and particle
production occurs when $|\dot m_\chi| \sim g|\dot \phi|$ becomes
greater than $m^2_\chi = g^2\phi^2$. For a harmonic oscillator one
has  $|\dot\phi_0| = m\Phi$, where $|\dot\phi_0|$ is the velocity
of the field  in the minimum of the effective potential, and $\Phi
\sim 10^{-1} M_p$ is the amplitude of the first oscillation. This
implies that the process becomes nonadiabatic for $g \phi^2
\lesssim m\Phi$, i.e.
for $ -\phi_* \lesssim  \phi \lesssim \phi_*$, where $\phi_* \sim
\sqrt{m\Phi\over g}$ \cite{KLS}. Here $\Phi \sim 10^{-1} M_p$ is
the initial amplitude of the oscillations of the inflaton field.
Note that under the condition $g\gg 10^{-4}$ which is necessary
for efficient reheating, the interval  $ -\phi_* \lesssim  \phi
\lesssim \phi_*$ is very narrow: $\phi_* \ll \Phi$. As a result,
the process of particle production occurs nearly instantaneously,
within the time
\begin{equation}\label{time}
\Delta t_* \sim  {\phi_*\over |\dot\phi_0|} \sim (g m \Phi)^{-1/2} .
\end{equation}
This time interval is much smaller than the age of the universe,
so all effects related to the expansion of the universe can be
neglected during the process of particle production. The
uncertainty principle implies in this case that the created
particles will have typical momenta $k \sim   (\Delta t_*)^{-1}
\sim  (g m \Phi)^{1/2}$. The occupation number  $n_k$ of $\chi$
particles with momentum $k$ is equal to zero
 all the time when it moves toward $\phi = 0$. When it reaches $\phi
= 0$ (or,
more exactly, after it moves through the small region  $ -\phi_*
\lesssim  \phi
\lesssim \phi_*$)  the occupation number  suddenly (within the time
$\Delta
t_*$) acquires the value \cite{KLS}
\begin{equation}\label{number}
n_k = \exp\left(-{\pi k^2 \over  gm\Phi}\right)  ,
\end{equation}
and this value does not change until the field $\phi$ rolls to the
point $\phi = 0$ again.

A detailed description of this process including the derivation of
Eq. (\ref{number}) was given   in
  the second paper of Ref.
\cite{KLS}; see in particular Eq. (55) there.
 This  equation  (\ref{number}) can be written in a  more general
form.  First of all, the shape of the effective potential does not
play any role in its derivation. The essential point of the
derivation of Eq. (\ref{number}) is that $\chi$ particles are
produced in a small vicinity of the point $\phi = 0$, when
$\phi(t)$ can be represented as
 $\phi(t)\approx
\dot \phi_0 (t -t_0)$.  The only   thing which  one needs to know
is not $V(\phi)$, $m$ or $\Phi$,  but the velocity of the field
$\phi$ at the time when it passes the point $\phi = 0$. Therefore
one can replace  $ m\Phi$ by $|\dot \phi_0|$ in this equation.
Also, the same equation is valid for massive particles $\chi$ as
well, if one replaces $k^2$ by $k^2 + m_\chi^2$, where $m_\chi$ is  
the bare
mass of the particles $\chi$ at $\phi = 0$. (A similar result
is valid for fermions and for vector particles.)
 Therefore Eq. (\ref{number})  in a general case
 (for any $m_\chi$ and $V(\phi)$)
   can be written   as follows:
\begin{equation}\label{general}
n_k = \exp\left(-{\pi (k^2+ m_\chi^2) \over  g|\dot\phi_0|}\right)  .
\end{equation}
 This can be integrated to give the
density of $\chi$ particles
\begin{equation}\label{suppr}
 n_{\chi}   =  {1 \over 2\pi^2  }
\int\limits_0^{\infty} dk\,k^2 n_k =  {({g\dot\phi_0})^{3/2} \over
8\pi^3} \exp
 \left(-{\pi  m_\chi^2  \over  g|\dot\phi_0|}\right)   .
\end{equation}
Numerical investigation of inflation in the theory ${m^2\over
2}\phi^2$  with $m = 10^{-6} M_p$ gives $|\dot\phi_0| = 10^{-7}
M_p^2$, whereas in the theory ${\lambda\over 4}\phi^4$  with
$\lambda = 10^{-13}$ one has a somewhat smaller value
$|\dot\phi_0| = 6\times 10^{-9} M_p^2$.  This implies, in
particular, that  if one takes $g \sim 1$, then in the theory
${m^2\over 2}\phi^2$ there is no exponential suppression of
production of $\chi$ particles unless their mass is greater than
$m_\chi \sim 2 \times 10^{15} $ GeV. This agrees with a similar
conclusion obtained in  \cite{KLS,Kolb,heavy,chung}.

Let us now concentrate on the case $ m_\chi^2 \lesssim
g|\dot\phi_0|$, when the
number of produced particles is not exponentially suppressed. In this case
\begin{equation}
 n_{\chi}   \approx   {({g\dot\phi_0})^{3/2} \over 8\pi^3}  \ .
\end{equation}

According to Eq. (\ref{general}), a typical initial energy
(momentum) of each particle  $\chi$ at the moment of their
production is $\sim (g|\dot\phi||/\pi)^{1/2}$, so their total
energy density is
\begin{equation}\label{rho_0}
\rho_{\chi}   \sim   {({g\dot\phi_0})^2 \over 8\pi^{7/2}}   \ .
\end{equation}
The
ratio of this energy to the total energy density $\rho_\phi =
\dot\phi_0^2/2$ of the scalar field $\phi$ at this moment gives
\begin{equation}\label{ratio_0}
{\rho_{\chi} \over \rho_\phi}  \sim  5\times 10^{-3}\, {g^{2} }  .
\end{equation}
  This result is practically model-independent, given the
interaction term $-{\textstyle {1 \over 2}} g^2\phi^2\chi^2$. In
particular, it does not depend on the inflaton mass $m$ in the
theory ${m^2\over 2}\phi^2$. The same result can be obtained in
the theory ${\lambda\over 4}\phi^4$ independently of the value of
$\lambda$.

An interesting possibility appears if one has  $m_\chi^2 \sim
g|\dot\phi_0|$. Then the probability of production of such
particles is not exponentially suppressed during the first
oscillation, but it is exponentially suppressed during all
subsequent oscillations because $|\dot\phi|$ decreases due to the
expansion of
the universe,
and the condition $ m_\chi^2 \lesssim g|\dot\phi|$   becomes
violated. In this
case new particles
$\chi$ are not created. However, as we already explained,   these
new particles
may not even be
necessary.  For example, in the theory ${\lambda\over 4}\phi^4$
the energy density of the inflaton field $\rho_\phi$ decreases as
$a^{-4}$, whereas the energy density stored in the nonrelativistic
particles $\chi$ (prior to their decay) decreases only as
$a^{-3}$. Therefore their energy density rapidly becomes dominant
even if originally it was small. Their subsequent decay makes the
process of reheating complete.

But preheating in our model becomes much  more efficient if we use the
mechanism
described in the beginning of this paper.
Indeed,  let us assume that the particles $\chi$ survive until the
field $\phi$
rolls up from $\phi = 0$ to the point $\phi_1$ from which it returns
back to
$\phi = 0$. In the theory ${m^2\over 2}\phi^2$ one has $ \phi_1
\approx  0.07
\,M_p$, whereas in the theory ${\lambda\over 4}\phi^4$ one has
$\phi_1 \approx
0.12\,M_p$. We will take  $ \phi_1
\approx  0.07
\,M_p$ in our
estimates. At
that time the mass of each particle $\chi$ will be  $g \phi_1  \sim
10^{-1}  g
M_p$, they will be nonrelativistic, and their total energy density
(for the
case of the theory ${m^2\over 2}\phi^2$) will be
\begin{equation}\label{rho}
\rho_{\chi}   =   m_\chi n_\chi \approx 10^{-1} g M_p
{({g\dot\phi_0})^{3/2}
\over 8\pi^3}  \sim   10^{-14} g^{5/2} M_p^4 \ .
\end{equation}
Therefore the ratio of the energy density of $\chi$ particles   to
the energy
density of the inflaton field $\sim \dot\phi_0^2/2$ will be
\begin{equation}\label{ratio}
{\rho_{\chi} \over \rho_\phi}  \sim   10^{-3} { |\dot\phi_0|} ^{-1/2} M_p
g^{5/2} \sim 2\, g^{5/2}  .
\end{equation}
  The last result follows from the relation $|\dot\phi_0|
\sim 10^{-1} m M_p \sim 10^{-7} M_p^2 $ for $m \sim 10^{-6} M_p$.
Under the condition $g \gtrsim 10^{-4}$, which is the standard
condition for efficient preheating \cite{KLS}, this ratio is much
greater than the one in Eq. (\ref{ratio_0}).

If the particles $\chi$ do not decay when the field $\phi$ reaches
$\phi_1$, then their energy will decrease again in parallel with
$|\phi|$, until it reaches the value given by Eq. (\ref{ratio_0}).
Thus,  preheating is most efficient if all particles $\chi$ can
decay at the moment when the field $\phi$ reaches its maximal
value $\phi_1$. This  is   possible if the lifetime of the
particles $\chi$ created  at the   moment $t_0$  is close to
$\Delta t \sim {\pi m^{-1}/4}$. Particles $\chi$ in our model can
decay to fermions, with the  decay rate \cite{KLS}
\begin{equation}\label{7}
\Gamma( \chi \to \psi \psi ) = { h^2 m_\chi\over 8 \pi} = { h^2 g
|\phi|\over 8
\pi}\ .
\end{equation}
Note that the decay rate grows with the growth of the field
$|\phi|$, so particles tend to decay at  large $|\phi|$. One can
easily check that the particles $\chi$   decay when the field
$\phi$ reaches its maximal value $|\phi| \approx 0.07 M_p$
if
\begin{equation}\label{8}
h^2 g   \sim {500 m\over M_p}  \sim 5\times10^{-4}.
\end{equation}
At the moment when $|\phi|$ reaches $0.07 M_p$, the particles
$\chi$ have effective mass $m_\chi = g|\phi| \sim 0.07 g M_p$.
Such particles can decay to two fermions $\psi$ if $ m_\psi <
0.035\, g M_p$. This implies that after the first half of an
oscillation, the scalar field $\phi$ can produce fermions with
mass up to $0.035\, g M_p$.  For example, in the theory with  $g
\sim 10^{-1}$, $h \sim 7\times 10^{-2}$  one can produce fermions
with mass up to $ m_\psi \sim  4 \times 10^{16}$ GeV, and in the
theory with  $g \sim  1$, $h   \sim 2 \times 10^{-2}$  one can
produce particles with mass up to $4 \times 10^{17}$ GeV.

As we have found, initially the ratio ${\rho_{\chi} \over
\rho_\phi}$ is suppressed by the factor $2 g^{5/2} $, see Eq.
(\ref{ratio}).
But
this suppression is not very strong, and if the energy density of
the $\psi$ particles during some short period of the evolution of
the universe decreases not as fast as the energy density  of the
inflaton field and other products of its subsequent decay, then
very soon the universe will be dominated by the products of decay
of the particles $\psi$, and reheating will  be complete.

 If $h^2 g \gg 5\times10^{-4}$, the $\chi$ particles may decay
before the oscillating field $\phi$ reaches its maximal value $\phi_1 \sim
10^{-1} M_p$. This can make our mechanism somewhat less efficient.
However,
the decay cannot occur until $m_\chi = g|\phi|$ becomes greater than $2
m_\psi$. If, for example, the fermions  have mass $\sim 0.03\, g
M_p$, then the
decay  occurs  only when the field $\phi$ reaches its maximal value
$\phi_1$
even if $h^2 g \gg 5\times10^{-4}$. This preserves the efficiency of our
mechanism even for very large $h^2 g$.

 On the other hand, for $h^2 g \ll 5\times10^{-4}$, the particles
$\chi$   do
not decay within a single oscillation. In this case the parametric
resonance
regime becomes possible, which again leads to efficient preheating
according to
\cite{KLS}. Moreover, superheavy fermions still will be produced in this
regime, because the oscillating field will spend a certain amount of  
time at
$\phi \sim \phi_1$. During this time superheavy particles will be
produced, and
their number may not be strongly suppressed.

The mechanism of particle production described above can work in a
broad class of theories. For example, one can consider models with
the interaction ${g^2\over 2}\chi^2(\phi + v)^2$. Such interaction
terms appear, for example, in supersymmetric models with
superpotentials of the type  $W = g \chi^2(\phi + v)$
\cite{berera}. In such models the mass $m_\chi$ vanishes not at
$\phi_1 = 0$, but at $\phi_1 = -v$, where $v$ can take any value.
Correspondingly,  the production of $\chi$ particles occurs not at
$\phi = 0$ but at $\phi = -v$. When the inflaton field reaches the
minimum of its effective potential at $\phi = 0$,  one has $m_\chi
\sim g v$, which may be very large. If one takes $v \sim M_p$, one
can get $m_\chi \sim g M_p$, which may be as great as $10^{18}$
GeV for $g \sim 10^{-1}$, or  even $ 10^{19}$ GeV for $g \sim 1$.
If, however, one takes $v \gg M_p$, the density of $\chi$
particles produced by this mechanism will be exponentially
suppressed by the subsequent stage of inflation. This possibility
will be discussed in the next section.

Since  parametric amplification of particle production is not
important in the context of the instant preheating scenario, it
will  work equally well if the inflaton field couples not to
bosons but to fermions \cite{baacke,GK98}. Indeed, the
  creation
of fermions with mass $g|\phi|$  also occurs because of the
nonadiabaticity of the change of their mass at   $ \phi  = 0$. The
theory of
this effect  at $g \gtrsim 10^{-4}$
is very similar to  the theory of the creation of $\chi$ particles
described above; see in this respect \cite{GK98}.   The efficiency
of preheating will be enhanced if the fermions $\psi$ with a
growing mass $g|\phi|$ can decay into other fermions and bosons,
as in the scenario described in the previous section.

It is   amazing that oscillations of the field $\phi$ with mass $m
= 10^{13}$ GeV can  lead to the copious production of superheavy
particles with masses 4 - 5 orders of magnitude greater than $m$.
 The previously known mechanism of preheating was
barely capable of producing particles of mass $\sim 10^{15}$ GeV,
which is somewhat below the GUT scale, and even that was possible
only in the strong coupling limit $g = O(1)$. Our new mechanism
allows for the production of particles with mass greater than
$10^{16}$ GeV even if the coupling constants are relatively small.
This fact may play an important role in the theory of baryogenesis
in GUTs.

\section{Fat wimpzillas}

  Until now we have discussed a new mechanism  of preheating.
However, recently there has been a growing interest in the
possibility of the production of superheavy WIMPS after inflation
\cite{KR,grav,wimppreh,chung}. Such particles (which have been
proudly called WIMPZILLAS \cite{WIMPZ}) could be responsible for the dark
matter content of the universe,   and, if they have very large but
finite decay time, they can also be responsible for cosmic rays
with energies greater than the Greisen--Zatsepin--Kuzmin limit
\cite{gzk}. The focus of these works in a certain sense was
opposite to that of the theory of preheating: It was necessary to
find a mechanism for the production of stable (or nearly stable)
particles which would survive until now. For that purpose, the
mechanism of their production must be extremely {\it inefficient},
since otherwise the present density of such relics would be
unacceptably large.

As one could expect, it is much easier to make the mechanism
inefficient rather
than the other way around. For example,  our Eq. (\ref{suppr})
implies that the
probability of production of superheavy $\chi$ particles is
suppressed by a
factor of $\exp  \left(-{\pi  m_\chi^2  \over
g|\dot\phi_0|}\right)$. In the
theory   ${m^2\over 2}\phi^2$  with $m = 10^{-6} M_p$ we have
$|\dot\phi_0| =
10^{-7} M_p^2$, so this suppression factor is given by $\exp
\left(-{10^7 \pi
m_\chi^2  \over  gM_p^2}\right)$.  This
implies that for $g \sim 1$ the production of particles with
$m_\chi \sim 10^{16}$ GeV is suppressed approximately by
$10^{-10}$, and this suppression becomes as strong as $10^{-40}$
for $m_\chi = 2\times 10^{16}$ GeV. This same level of suppression
can be achieved, for example, with $g = 10^{-2}$ and $m_\chi =
2\times 10^{15}$. Thus, by fine-tuning of the parameters
$m_\chi$ and $g$ one can obtain any value of the density of WIMPS
at the present stage of the evolution of the universe. This result
agrees with the result obtained in \cite{chung} by a different
method.

This suppression mechanism is equally operative for the process
$\phi \to \chi \to \psi$ discussed in our paper.  If the particles
$\chi$ are heavy at $\phi = 0$, their number will be exponentially
suppressed. When the field $\phi$ grows, their masses grow as
follows: $m_\chi^2(\phi) = m_\chi^2 + g^2\phi^2$, At the moment of
their decay these particles can have mass of the order $10^{17} -
10^{18}$ GeV. The main advantage of this new mechanism is that the
process of fattening of the particles $\chi$ described above
allows for the production of particles $\psi$  which can be $10^2$
times heavier than their   cousins discussed in
\cite{KR,grav,wimppreh,chung}. In the absence of established
terminology, one can call such superheavy particles FAT
WIMPZILLAS.

Another way to produce an exponentially small number of superheavy
WIMPS is to produce them at the last stages of inflation. This is
possible in theories with the interaction term ${g^2\over
2}\chi^2(\phi+ v)^2$, as described in the previous section. If one
takes $v \gtrsim M_p$, then the particles $\chi$
 will be
created during inflation.  The number of $\chi$ particles
produced during inflation in the simplest theory with  $V(\phi)
= {m^2\over 2} \phi^2$ does not depend on $v$ because $\dot
\phi$ does not depend on $\phi$ and on $v$ in this scenario:  $\dot \phi=
{mM_p\over 2\pi}$ \cite{book}.  However, their density
will subsequently be exponentially suppressed by inflation. This
is exactly what we need
if the $\chi$ particles or the products of their decay are WIMPS.
For example, in the theory with $V(\phi) = {m^2\over 2} \phi^2$
the universe inflates  by a factor of $\exp ({2\pi v^2/M_p^2})$
after the creation of $\chi$ particles \cite{book}, so their
density at the end of inflation becomes smaller by a factor of
$\exp ({6\pi v^2/M_p^2})$. This leads to a desirable suppression
for $v \sim 2 M_p$. (The exact number depends on the subsequent
thermal history of the universe.) Meanwhile the masses of WIMPS
produced by this mechanism can be extremely large, of order
$gM_p$. If the $\chi$ particles are stable, they themselves may
serve as superheavy WIMPS with nearly Planckian mass. If they
decay to fermions, then the fermions may play a similar role.

\section{Quintessence, instant preheating,  and black hole production}

The mechanism of instant preheating   works even better in models
with potentials of a ``quintessential'' type. For example, one may
consider potentials $V(\phi)$ which behave (approximately) as $
{m^2\over 2} \phi^2$ or ${\lambda\over 4}\phi^4$ at $\phi < 0$,
and (gradually) vanish when  $\phi $ becomes positive. Models of
this type were first proposed by Ford and Spokoiny, and recently
they have been revived by several other authors \cite{spok}. These
models for a long time were  unpopular because the field $\phi$ in
these models does not oscillate and therefore the old mechanism of
reheating based on the perturbative decay of the inflaton field
\cite{DL} in such models cannot work. The only known way to reheat
the universe  in this scenario was to invoke the gravitational
creation of particles due to the expansion of the universe
\cite{grib,spok}. This mechanism of reheating is relatively
inefficient and may lead to several cosmological problems which we
will discuss in a separate publication  \cite{fkls}. Fortunately,  
our mechanism
works in these theories as well. Indeed, the theory of production
of particles $\chi$ in our scenario is applicable to theories with
any $V(\phi)$; the number of produced particles depends only on
$g$ and $|\dot \phi_0|$. In  the theories where $V(\phi)$
gradually vanishes
at large $\phi$,  the field $\phi$ rolls at least up to $ \phi \sim
M_p$ before
it stops (or
turns back). Then the particles $\chi$ acquire masses $m_\chi \sim
g M_p $. If the coupling constant $g$ is sufficiently large ($g
\sim 10^{-1}$), the decay of $\chi$ particles may
produce superheavy particles $\psi$ with masses up to $10^{17} -
10^{18} $ GeV. Since the value of the field $\phi$ at that time is
approximately one order of magnitude greater than in the usual
model with the potential $ {m^2\over 2} \phi^2$, whereas the energy  
density
$\rho_\phi$ will be decrease due to the expansion of the universe,  
the energy
density of particles $\chi$ at that time will be much
greater than $ 2 g^{5/2} \rho_\phi$.

It is instructive to compare the density of particles produced by
this mechanism  to the density of particles created during
gravitational particle production, which is given by $\rho_\chi
\lesssim O(H^4) \sim \rho_\phi {\rho_\phi\over M_p^4}$. In the
simplest models of chaotic inflation ${\rho_{\chi} \over
\rho_\phi} \sim {\rho_\phi\over M_p^4} \sim 10^{-14}$ at the end
of inflation, and in hybrid inflation models it is even much
smaller. Thus, for $g \gtrsim 10^{-6}$ the number of particles
produced during instant preheating  is much greater than the
number of particles produced by gravitational effects. Therefore
one may argue that reheating of the universe in theories with
quintessential potentials should be described using
  the instant preheating scenario. As for the gravitational particle
production studied in \cite{spok}, it still can be useful, perhaps
not as a mechanism of reheating, but as a source of WIMPS. Indeed,
as we already mentioned, the number of WIMPS must be extremely
small, so the relative inefficiency of the gravitational particle
production is quite appropriate in this context \cite{grav}.

In theories with  quintessential potentials  the energy density of
the inflaton field $\phi$ rolling along the flat direction of the
effective potential decreases as $a^{-6}$, i.e. {\it much} faster
than the energy density of other particles. Thus, in such theories
instant preheating   is amazingly
efficient. Even if the particles $\chi$ decay to relativistic
particles immediately after they are produced, so that the
fraction of energy transferred to particles from the inflaton
field initially   is only $5\times 10^{-3} g^{2}$,  still their energy
density will dominate as soon as the size of the universe grows by
a factor of   $15 g^{-1}$.   Meanwhile, if one uses our favorite  
mechanism   by
which the $\chi$ particles ``fatten" before decaying, then
the energy density of produced particles becomes
much greater, and the period of their dominance with respect to
the energy density of the classical inflaton field begins even
much earlier,  when the scale factor grows by only a factor of  
$O(g^{-5/6})$.

Note that in this scenario   we still assume that   $\chi$ particles  
eventually
decay to some other particles. Otherwise the
backreaction of the created $\chi$ particles can stop the growth
of the field $\phi$ and return it back to $\phi = 0$.
However,
since the number of produced particles and their interaction with
the field $\phi$ are suppressed for small $g$,   the
backreaction of the created particles becomes significant only
after a time interval much greater than $m^{-1}$  \cite{fkls}.  As a  
result,
this scenario will work under much milder constraints on $g$ and
$h$ than the scenarios in which $\phi$ oscillates about a minimum
of $V(\phi)$.

One can avoid this problem altogether if
$V(\phi)$ becomes  flat not at $\phi = 0$, but only  at   $\phi
\gtrsim M_p$. In such a case the backreaction of created particles
never turns the scalar field $\phi$ back to $\phi = 0$. Therefore
the decay of the particles $\chi$ may occur very late, and one can
have efficient preheating for any values of the coupling constants
$g$ and $h$.

One may also consider a scenario in which the particles $\chi$ are
stable, and have bare mass $m_\chi \sim   10^{16}$ GeV. The
probability of production of such particles will be exponentially
suppressed, but then their masses will increase by $g|\phi|$,
which can be very large. This  allows to produce superheavy
WIMPS of mass $m_\chi \gg 10^{16}$ GeV without assuming that
$\chi$ particles must decay to $\psi$ (see also the previous section).

Finally, let us assume that the $\chi$ particles are stable, and
have bare mass $m_\chi \ll  10^{16}$ GeV. Then the probability of
creation of such particles will be large, and if they cannot decay
to other particles, we will eventually end up with the universe
filled by an unacceptably large number of superheavy WIMPS.
However, if the field $\phi$ continues rolling for a very long
time, it may reach values much greater than $M_p$. In such a
scenario, the particle masses $g|\phi|$  at some moment may become
even larger than $M_p$.

The possibility of producing superheavy
particles with masses exceeding $M_p$ should be addressed in the
framework of  superstring theory. It is rather interesting that
superstring theory in certain cases may be important for the
description of
reheating after inflation, which  was viewed as a low-energy
phenomenon.

In conventional  quantum field theory, an elementary particle of
mass $M$ has a Compton wavelength $M^{-1}$ smaller than its
Schwarzschild radius $2 M/M_p^2$ if $M > M_p$. Therefore one may
expect that as soon as   $m_\chi = g|\phi|$ becomes greater than
$M_p$, each $\chi$ particle  becomes a Planck-size black hole,
which immediately evaporates and reheats the universe. This is a
very unusual (and admittedly very speculative) version of the
instant preheating scenario which deserves more detailed
investigation.

\section{Conclusions}
The theory of reheating after inflation is already rather old. For
many years we thought that the classical oscillating inflaton
field could be represented as a collection of scalar particles of
mass $m \lesssim 10^{13}$ GeV, that each particle decayed to
particles of smaller mass, and that our final goal was to
calculate the reheating temperature $T_r$.

During the last few years we have learned that   this simple
picture in certain cases  can be very useful, but typically
one must use the nonperturbative theory of reheating for the
description of the first stages of reheating. The main ingredient
of this theory was the theory of broad parametric resonance.
Particle production in this scenario could be represented as a
series of successive acts of creation, during which the number of
produced particles increased exponentially. It seems now that this was
only a first step towards a complete understanding of
nonperturbative mechanisms of reheating after inflation. As we
have shown in this paper, preheating may occur in a different way.
It may be sufficient to consider a single act of creation,
especially if one takes into account the relative increase
  of energy of
produced particles during the subsequent evolution of the
classical inflaton field, and the possibility of the chain
reaction $\phi \to \chi\to \psi$. This new mechanism is capable of
producing particles of  nearly Planckian energy, which was
impossible in the previous versions of the theory of reheating.

One of the key ingredients of the nonperturbative mechanism of preheating
described above is a  nonadiabatic change of the mass $m_\chi(\phi)$  
near the
point   where it vanishes (or at least strongly decreases).  Such  
situations
occur very naturally  in supersymmetric theories of elementary  
particles if one
identifies $\phi$ and $\chi$ with moduli fields which correspond to flat
directions of the effective potential. Indeed, in supersymmetric  
theories   the
effective potential often has several flat directions, which may  
intersect.
When one of the moduli fields (the inflaton) moves along a flat  
direction and
reaches the intersection, the mass of another field vanishes. A simplest
example of this situation was described in Sect. III. The change of  
the number
of massless degrees of freedom is a generic phenomenon which is  
under intense
investigation in the context of  supersymmetric gauge theories,  
supergravity
and string theory, where it is associated with the points of  
enhanced gauge
symmetry, see e.g. \cite{sieberg,polch}.

Masses of elementary particles may also change nonadiabatically during
cosmological phase transitions. At the moment of a phase transition  
masses of
some particles vanish and   may even temporarily become tachyonic.  
In this case
particle production may become even more intense.

Our main conclusion is that with an account taken of the new possibilities
discussed above the scenario of preheating becomes more robust. In the
cases where parametric resonance may occur, it provides a very
efficient mechanism of preheating. Now we have found that
efficient preheating is possible even  in models where parametric
resonance does not happen because of the rapid decay of produced
particles.  Instant preheating occurs in the usual inflationary
models where the inflaton field oscillates near the minimum of its
effective potential. But this mechanism works especially well in
models with effective potentials which slowly decrease at large
$\phi$, as in
the theory of quintessence.
 The conversion of the
energy of the inflaton field to the energy of elementary particles
in these models occurs very rapidly, and it is always 100\%
  efficient. A preliminary investigation
indicates that in some versions of such models preheating may even
produce particles of mass greater than $M_p$ which   become black
holes and immediately evaporate. It would be very interesting to
investigate this possibility in the context of string theory.

\bigskip
\section*{Acknowledgments}
It is a pleasure to thank R. Kallosh, S. Shenker,  A. Starobinsky,  
S. Thomas,
and A. Vilenkin for
useful  discussions.  This work was supported  by NSF grant AST95-29-225.
The work of A.L.  was also supported   by NSF grant PHY-9870115.


\begin{references}
\bibitem{super} Note that if one takes $g > 10^{-3}$,   radiative  
corrections
to the effective potential may considerably change its shape \cite{book}.
However, this does not happen in supersymmetric theories where the
contributions of fermions and bosons nearly cancel each other.
\end{references}
\end{document}